\documentstyle[epsf]{mn}
\loadboldmathitalic 
\newcommand{\mm}[1]{\mbox{$#1$}} 
\newcommand{\ltsim}{\la}
\newcommand{\gtsim}{\ga}
\newcommand{\unit}[1]{\ifmmode \:\mbox{\rm #1}\else \mbox{#1}\fi} 
\newcommand{\bv}[1]{{\bmi #1}} 
\renewcommand{\sb}[1]{_{\rm #1}} 
\newcommand{\mone}{\mm{^{-1}}} 
\newcommand{\dif}{\mbox{d}} 
 
\newcommand{\kms}{\unit{km~s\mone}} 
 
\newcommand{\mpc}{\unit{Mpc}} 
 
\newcommand{\hmpc}{\mm{h\mone}\mpc} 
 
\newcommand{\dnsig}{\mm{D\sb{n}-\sigma}} 
\newcommand{\om}{\mm{\Omega}}

\newcommand{\dgr}{\degr} 
\newcommand{\ddgr}{\fdg} 
 
\newcommand{\dec}{\mbox{Dec.}} 
\newcommand{\secref}[1]{Section~\ref{sec:#1}} 
 
\newcommand{\eqref}[1]{equation~(\ref{eq:#1})} 
 
\newcommand{\figref}[1]{Fig.~\ref{fig:#1}} 
\newcommand{\tabref}[1]{Table~\ref{tab:#1}} 
\newcommand{\sbo}{\sb{O}} 
\newcommand{\sigd}{\sigma_{\delta}} 
\newcommand{\sigadd}{\sigma\sb{add}} 
\newcommand{\POTENT}{{\sc potent}}
\title 
[Optical galaxies versus \POTENT\ mass] 
{$\om$ and biasing from optical galaxies versus \POTENT\ mass} 
\author[M. J. Hudson et al.] 
{M.J. Hudson,$^{1}$ A. Dekel,$^{2}$ 
S. Courteau,$^{3}$ S.M. Faber$^{4}$ and J.A. Willick$^{5}$ \\ 
$^{1}$ Dept.\ of Physics, Univ.\ of Durham, South Road, Durham 
DH1 3LE. 
E-mail: M.J.Hudson@durham.ac.uk\\ 
$^{2}$ Racah Institute of Physics, 
The Hebrew University of Jerusalem, Jerusalem 91904, Israel. 
E-mail: dekel@vms.huji.ac.il\\ 
$^{3}$ Kitt Peak National Observatory, P.O. Box 26732, 
Tucson, AZ 85726-6732, USA. E-mail: courteau@noao.edu \\ 
$^{4}$ UCO/Lick Observatory, University of California, Santa Cruz, 
CA 95064, USA. E-mail: faber@lick.ucsc.edu\\ 
$^{5}$ Carnegie Observatories, 
813 Santa Barbara St., Pasadena, CA 91101, USA. 
E-mail: jeffw@cetus.ociw.edu 
} 
\date{Accepted 1995 January 4. Received 1994 December 29; in original
form 1994 October 26}
\begin{document} 
\maketitle 
\begin{abstract} 

The mass density field in the local universe, recovered by the \POTENT\
method from peculiar velocities of $\sim$3000 galaxies, is compared
with the density field of optically selected galaxies.  Both density
fields are smoothed with a Gaussian filter of radius 12$h^{-1}$ Mpc.
Under the assumptions of gravitational instability and a linear
biasing parameter $b\sbo$ between optical galaxies and mass, we obtain
$\beta\sbo \equiv \om^{0.6}/b\sbo = 0.74 \pm 0.13$.  This result is
obtained from a regression of \POTENT\ mass density on optical density
after correcting the mass density field for systematic biases in the
velocity data and \POTENT\ method.  The error quoted is just the
$1\sigma$ formal error estimated from the observed scatter in the
density--density scatterplot; it does not include the uncertainty due
to cosmic scatter in the mean density or in the biasing relation. We
do not attempt a formal analysis of the goodness of fit, but the
scatter about the fit is consistent with our estimates of the
uncertainties.

\end{abstract} 


\section{Introduction} 
\label{sec:intro} 

With the advent of large-scale redshift and peculiar velocity surveys, 
it has become possible to investigate the relationship between the 
distribution of mass and galaxies on supercluster scales. Assuming 
that structures grow through gravitational instability (GI), then, on 
large scales where linear theory is applicable, the peculiar velocity 
of a galaxy is proportional to its peculiar gravitational 
acceleration. 
The constant of proportionality depends on the 
density parameter, $\om$, like 
$f(\om) \simeq \om^{0.6}$ (e.g.\ Peebles 1980). 
The acceleration scales with the 
amplitude of the 
mass density fluctuation field, so,
if we assume that the density contrast of 
optically selected galaxies, $\delta\sbo$, is related to the density 
contrast of mass, $\delta$, by linear biasing 
$\delta\sbo = b\sbo \delta$, then the appropriate scaling factor 
between 
peculiar 
velocity and galaxy density 
becomes $\beta\sbo \equiv f(\om)/b\sbo$. 
The determination of $\beta\sbo$ is the goal of this paper. 

There are two general ways 
to obtain $\beta\sbo$ from 
peculiar velocity data and the density field 
as derived from a galaxy redshift survey. 
Each method has advantages and disadvantages. 
One approach is to predict radial peculiar velocities from the density 
field via gravity and adjust $\beta\sbo$ for the best fit to the 
peculiar velocity data 
(e.g. Huchra 1988 and references therein; 
Strauss 1989; Kaiser et al.\ 1991; Shaya, Tully \& Pierce 
1992; Hudson 1994b; Willick et al., in preparation). 
A potential error in such velocity--velocity comparisons arises 
from the non-locality of the predicted velocities, 
which are affected by the density field in unsurveyed regions 
such as the Galactic Zone of Avoidance and distant regions. 
Another potential source of error is the fact that 
the computation of the predicted velocities involves smoothing,
while they are compared to raw, unsmoothed peculiar velocity data. 

An alternative approach is to compare density fields smoothed in a 
similar way. In linear theory, the mass density fluctuation can be 
determined from the divergence of the peculiar velocity field, 
$\bv{v}$, 
\begin{equation} 
\delta = - f^{-1} \bv{\nabla} \cdot \bv{v} 
\ , 
\end{equation} 
and a generalized approximation is valid in the quasi-linear regime 
(see \secref{pot} below). The comparison of densities in the linear 
regime yields $\beta\sbo$, and in the quasi-linear regime it allows a 
determination of $b\sbo$ for a given value of $\om$. The main 
advantages of this approach are the locality of the comparison 
(on scales larger than 
the smoothing scale), and the fact that similar smoothing is applied 
to {\em both \/} fields at a scale on which linear or quasi-linear 
theory is applicable. This approach was taken by Dekel et al.\ (1993, 
hereafter DBY93), who compared the mass density derived by \POTENT\ 
(Dekel, Bertschinger \& Faber 1990, hereafter DBF; Bertschinger et al.\ 
1990) from the peculiar velocity sample of Burstein (1990, Mark II) to 
the density field of galaxies drawn from the {\it IRAS\/} 1.9-Jy redshift 
survey (Strauss et al.\ 1990, 1992). 
Dekel et al.\ used 
an elaborate likelihood analysis 
to find 
that the data were consistent with the assumptions 
of GI and linear biasing, and estimated $\beta_I=1.3\pm0.3$. 

The goal of this paper is to estimate $\beta\sbo$, under the 
assumption of GI and linear biasing, by a straightforward comparison 
of the density field of {\em optically selected\/} galaxies and the 
mass density field as determined by an improved \POTENT\ procedure 
applied to an extended and newly calibrated data set. 

The outline of this paper is as follows. The mass density field is 
first determined by applying \POTENT\ to a data set of peculiar 
velocities containing $\sim$3000 galaxies. The velocity data, some 
details of the \POTENT\ method, and in particular the biases introduced 
in using sparse, non-uniform and noisy radial peculiar velocities to 
obtain a mass density field are summarized in \secref{pot}. The 
optical density field is then described in \secref{opt}. 
In \secref{comp}, we estimate $\beta\sbo$ by means of a linear regression 
of the \POTENT\ density fluctuation, $\delta\sb{P}$, on the optical 
density fluctuation, $\delta\sbo$, eliminating biases in the \POTENT\ 
procedure, and evaluate the robustness of the result. 
We also demonstrate the difficulty 
of separating 
$\om$ and $b\sbo$ via non-linear effects. Finally, 
in \secref{discuss}, we summarize our results and discuss them in 
light of other determinations.

\section{POTENT density field} 
\label{sec:pot} 

\def\vx{\bv{x}} \def\vr{\bv{r}} \def\vv{\bv{v}} 
\def\vs{\bv{s}} \def\ve{\bv{e}} 
\def\vB{\bv{B}} \def\vL{\bv{L}} 

The sample of galaxies with measured redshifts and forward 
Tully-Fisher or \dnsig\ (hereafter collectively TF) inferred distances 
used in this paper 
is a preliminary version of the Mark III catalogue. 
The calibration and compilation of the several data sets 
comprising the Mark III catalogue are described in detail by 
Willick et al.\ (1995 and in preparation). The elliptical data (Lucey \& 
Carter 1988; Faber et al.\ 1989; Dressler, Faber \& Burstein 1991) are 
the same data previously compiled by Burstein (1990) 
for 
the Mark II catalogue. 
The spiral data contain the nearby Mark II data of Aaronson 
et al.\ (1982) as revised by Tormen \& Burstein (1995), and newer 
extended data from Han (1991), Han \& Mould (1992), Mould et al.\ (1991, 
1993), Willick (1991), Courteau (1992), 
and Mathewson, Ford \& Buchorn (1992). The 
sample 
consists of 
about 3000 galaxies grouped into $\sim$1150 
objects. 

Inhomogeneous Malmquist (IM) bias affects the determination of 
$\om$ (or $\beta$) 
from TF data when the analysis, as in \POTENT, combines neighbouring 
galaxies in inferred-distance space (e.g. Willick 1991, 1994; 
Hudson 1994a; Dekel 1994). The distance errors, combined with galaxy-density 
variations along the line of sight, systematically enhance the 
inferred velocity gradients and thus the inferred density fluctuations 
and $\om$ 
or 
$\beta$. 
The IM bias is 
first 
reduced 
and
then 
corrected 
by a 
new 
procedure 
that has been tested using mock 
peculiar velocity 
catalogues 
drawn 
from N-body simulations. 
The mock catalogues are constructed in a way which mimics the 
non-uniform sampling and the noise of the real data. 
The IM bias correction procedure is 
discussed in detail by Dekel et al.\ (in preparation, hereafter D95c) 
and is only summarized briefly here. 
First, 
the galaxies are grouped heavily 
in redshift space, while correcting for `selection' bias 
(Willick 1991, 1994). 
This grouping reduces 
the distance error of each group of $N$ 
members to $\Delta/\sqrt N$, where $\Delta\approx 0.15-0.21$ and 
$0.21$ for spirals and ellipticals respectively. 
The IM bias scales like the square of the distance error, so this grouping 
significantly reduces the IM bias. Then, the noisy inferred distance 
of each object, $d$, is replaced by the expectation value of the true 
distance $r$ given $d$ (e.g. Willick 1991, equation 5.70): 
\begin{equation} 
E(r \vert d)= 
{ 
\int_0^\infty r^3 n(r)\ {\rm exp} 
\left( -{[{\rm ln}(r/d)]^2 \over 2\Delta^2} \right) {\rm d}r 
\over 
\int_0^\infty r^2 n(r)\ {\rm exp} 
\left( -{[{\rm ln}(r/d)]^2 \over 2\Delta^2} \right) {\rm d}r 
} \ . 
\end{equation} 
For single, ungrouped galaxies, $n(r)$ should be the number density of 
galaxies in the underlying distribution from which galaxies were selected 
for the sample (assuming they were selected 
by quantities that do not explicitly depend on $r$). 
In cases where the sample is redshift-limited, $n(r)$ is truncated at 
the appropriate distance. 
In a grouped sample, the density run $n(r)$ is multiplied by 
appropriate grouping/selection correction factors for the grouped and 
ungrouped objects respectively. 
(see D95c for details). 
This procedure 
is 
found to reduce the IM bias 
in the mock catalogues to the level of a few per cent. As an 
approximation to $n(r)$ we use for spirals the density of 
{\it IRAS\/} 1.2-Jy 
galaxies (Fisher 1992), and for ellipticals the density of 
early-type galaxies derived by Hudson (1993a, 
see also \secref{opt} below), both 
smoothed with a Gaussian filter of radius 
$500\kms$. 
The final IM correction typically amounts to less than 10 per cent in 
the $1200 \kms$ smoothed density field described next. 

The \POTENT\ procedure takes these discrete and noisy radial velocity 
data, $u_i$ at positions $\vx_i$, and first smoothes them with a 
Gaussian window of radius $1200 \kms$ (DBF; 
Dekel et al., in preparation, hereafter D95a, D95b).
It then applies the gravitational ansatz of potential flow to recover the 
missing transverse components of the velocity field (Bertschinger \& 
Dekel 1989). Finally, it applies a quasi-linear approximation with an 
assumed value of $\om$ to reconstruct the associated field of 
mass-density fluctuations. The quasi-linear generalization of the 
linear approximation to gravitational instability is the solution of 
the continuity equation under the Zel'dovich approximation in Eulerian 
space (Nusser et al.\ 1991), 
\begin{equation} 
\delta\sb{c}(\vx) = \Vert {\rm {\bf I}} - f^{-1} {\partial \vv / \partial \vx} 
\Vert -1 \ , 
\end{equation} 
where the bars denote the Jacobian determinant, and ${\rm {\bf I}}$ 
is the unit matrix. 

The non-trivial step in \POTENT\ is the {\it smoothing} of the data 
into a radial velocity field $u(\vx)$. This 
smoothing 
procedure is described and 
tested in detail elsewhere 
(DBF; D95b; Kolatt et al., in preparation). 
The procedure 
is only briefly summarized here 
with emphasis 
on the new features of the method. 
The aim is to reproduce the 
$u(\vx)$ that would have been obtained had the true three-dimensional 
velocity field $\vv(\vx)$ been sampled densely and uniformly and 
smoothed with a spherical Gaussian window of radius $R\sb{s}$. 
With the 
discrete 
data 
available, 
$u(\vx\sb{c})$ is taken to be the value at 
$\vx\!=\!\vx\sb{c}$ of an appropriate parametric local velocity model 
$\vv(\alpha_k,\vx\!-\!\vx\sb{c})$ obtained by minimizing the weighted sum 
of residuals 
\begin{equation} 
\sum_i W_i\, [u_i-\hat{\vx}_i\cdot\vv(\alpha_k,\vx_i-\vx\sb{c})]^2 
\ , 
\end{equation} 
in terms of the parameters $\alpha_k$ within an appropriate local window 
$W_i\!=\!W(\vx_i,\vx\sb{c})$. 
The local velocity model and the 
weighting needed in order to minimize random errors and systematic 
biases 
are chosen as follows.

Even for the case of dense, uniform sampling the recovered $u(\bv{x})$ 
field will in general suffer from `window bias'. 
Unless 
$R\sb{s}$ is much smaller than the distance from the origin to the centre 
of the window, $r\sb{c}$, the radial peculiar velocity data, $u_i$, cannot 
be averaged as scalars because the directions $\hat{\vx}_i$ differ 
from $\hat{\vx}\sb{c}$, so $u(\vx\sb{c})$ requires a fit of a 
local 3D velocity model. 
The original \POTENT\ 
method 
(DBF) used the simplest local model, $\vv(\vx)\!=\!\vB$, 
where $\vB$ is a uniform velocity,
for which the solution can be expressed explicitly in terms of a tensor 
window function. 
The tensorial correction to the spherical window has conical symmetry, 
weighting more heavily objects of large $\hat{\vx}_i\!\cdot\!\hat{\vx}\sb{c}$. 
This window deformation introduces a bias 
which 
is particularly severe nearby, where $R\sb{s}/r\sb{c}$ is not small, and 
in places 
where the velocity has a large divergence or convergence 
transverse to the line of sight. 
For example, a converging flow in a plane transverse to the line of sight 
would be biased to show an artificial component towards the origin. 
For the current peculiar velocity data, 
the resulting bias in the smoothed radial peculiar velocity field 
reaches 
a maximum value of 
$\sim\!300\kms$ at the 
position of the 
Great Attractor. A way to reduce this 
bias is by generalizing 
$\vv(\vx)$ 
into a {\it linear} velocity model, 
$\vv(\vx)\!=\!\vB\! +\! {\rm{\bf L}}\! \cdot\! (\vx\!-\!\vx\sb{c}) $, 
with ${\rm{\bf L}}$ a symmetric tensor, which ensures local 
irrotationality. The linear terms tend to `absorb' most of the 
bias, leaving $\vv(\vx\sb{c})\!=\!\vB$ less biased. Unfortunately, a 
high-order model tends to pick undesired small-scale noise, especially 
where the data are sparse and noisy. The optimal compromise is found 
to be a first-order model fit out to $r\!=\!40\hmpc$, smoothly 
changing to a zeroth-order fit beyond $60\hmpc$ 
(D95b).

Unfortunately, 
the real peculiar velocity data are both non-uniformly sampled 
and noisy. 
If the true velocity field is varying within the effective window, 
then 
non-uniform sampling 
will introduce a sampling-gradient (SG) 
bias. 
This bias occurs 
because the smoothing is weighted by the distribution of sampled 
galaxies whereas the aim is equal-volume weighting. One should weight 
each object by the local volume it `occupies', 
$V_i$. We adopt 
$V_i\!\propto\!R_4^3$ where $R_4$ is the distance to the fourth 
neighbouring object. 
This weighting procedure (termed P0) is found 
from 
simulations to 
reduce the SG bias in the Mark III sample to negligible levels 
out to 
a typical depth of 
$60\hmpc$ (away from the Galactic Zone of Avoidance). 
The $R_4(\vx)$ field 
will 
serve later as a flag for poorly sampled 
regions, to be excluded from any quantitative analysis. 

The ideal weighting for reducing the 
effect of Gaussian noise 
in the peculiar velocity measurements, 
$\sigma_i$, requires weights 
$W_i\!\propto\!\sigma_i^{-2}$, but this spoils the 
volume weighting 
and biases 
$u$ towards its values at smaller $r_i$ and at 
nearby clusters where the errors are small. A successful compromise 
(termed P1) is to weight by both, i.e. 
\begin{equation} 
W(\vx_i,\vx\sb{c}) \propto 
V_i\, \sigma_i^{-2}\, \exp [-(\vx_i-\vx\sb{c})^2/ 2 R\sb{s}^2] \ . 
\end{equation} 
The difference between the recovered fields using the P0 and P1 
weighting schemes can serve as an indicator for the magnitude of the 
systematic uncertainty still remaining in the \POTENT\ recovery. 

The errors in the recovered fields, due to scatter in the distance 
indicator and measurement errors, are assessed by Monte Carlo 
simulations, where the input distances are perturbed at random using a 
Gaussian of standard deviation $\sigma_i$ [in fact 
$(\sigma_i^2+\sigma\sb{f}^2)^{1/2}$, where $\sigma\sb{f}\sim 200\kms$ is the 
local dispersion of velocities in the `field'] before being fed into 
\POTENT. The error in $\delta\sb{P}$ at a grid point is estimated by 
the standard deviation of the recovered $\delta\sb{P}$ over the 
Monte Carlo simulations, $\sigma_{\delta}$. In the well-sampled 
regions, which extend in Mark III out to $40\!-\!60\hmpc$, the errors 
are $\sigma_{\delta}\!\approx\!0.1\! - \!0.3$, but they may blow up in 
certain regions at large distances. To exclude noisy regions, any 
quantitative analysis should be limited to points where 
$\sigma_{\delta}$ is within certain limits 
(see 
\secref{vol} 
below). 

There exists a fundamental freedom in determining the {\it zero-point} 
TF parameter which fixes the distance scale in units of $\kms$. 
A change of this zero-point multiplies all distances by a factor 
($1+\epsilon$) while the redshifts are fixed. This is equivalent to 
adding a monopole Hubble-like component $-\epsilon r$ to the peculiar 
velocities $\vv$, and an offset $3\epsilon f^{-1}(\om)$ to $\delta$ 
($\approx -f^{-1}\bv{\nabla}\cdot\vv$). The zero-point Hubble flow 
has tentatively been determined in the Mark III data from the 38 
spiral clusters, which are distributed across the the sample volume. 
After processing through \POTENT, the zero-point is fine-tuned by 
minimizing the volume-weighted variance of the recovered peculiar 
velocity field (equivalent in the linear regime to enforcing $\langle 
\delta \rangle=0$) in a `fair' volume, taken here to be a sphere of 
radius 6000$\kms$. 
In fact, the zero-point adopted has no effect on the determination 
of $\beta$ below 
(\secref{fit}). 

\begin{figure*}
\begin{center}
\epsfysize=22cm
\mbox{
\epsfbox{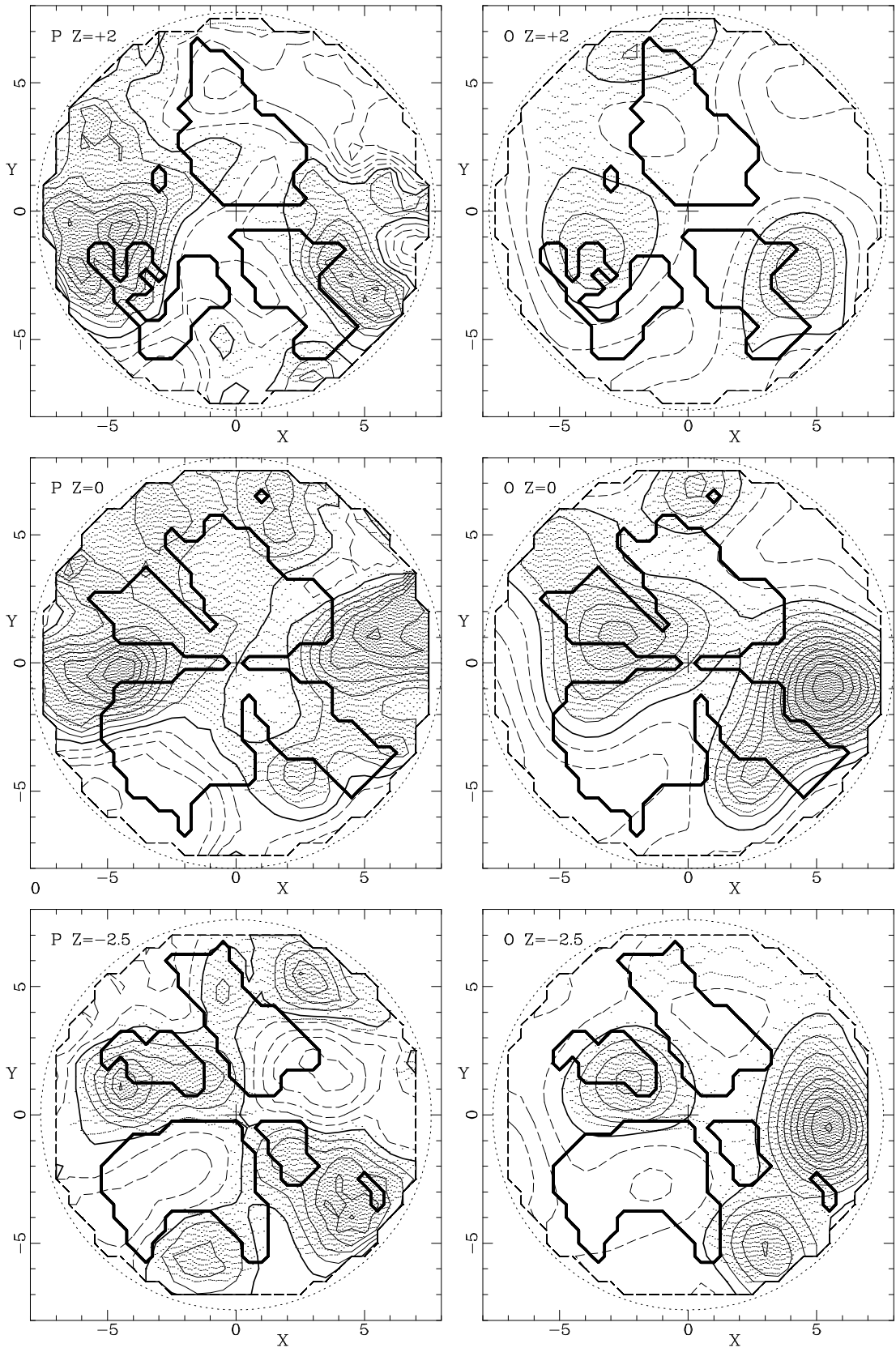}
}
\end{center}
\caption[]{
The density fields in slices parallel to the Supergalactic Plane.  
The left-hand panels show the \POTENT\ mass density field and the
right-hand panels show the density field of optical galaxies,
both smoothed with a Gaussian filter of radius 1200 \kms. 
The contour levels are in steps of $\Delta \delta = 0.2$.  The
medium-thickness contours
indicate the mean density,
thin solid contours show overdense regions
and dashed contours show underdense
regions.  The heavy contours indicate the boundary of the `Large'
comparison volume (see \secref{vol} for details).  
The dotted circle has a radius of 8000 \kms.
The top row shows
the slice at Supergalactic $Z=2000 \kms$, the middle row shows the
Supergalactic Plane and the bottom row shows the plane at
Supergalactic $Z=-2500 \kms$.  These slices are approximately independent.
}
\label{fig:maps}
\end{figure*}

\begin{figure*}
\begin{center}
\epsfysize=6cm
\mbox{
\epsfbox{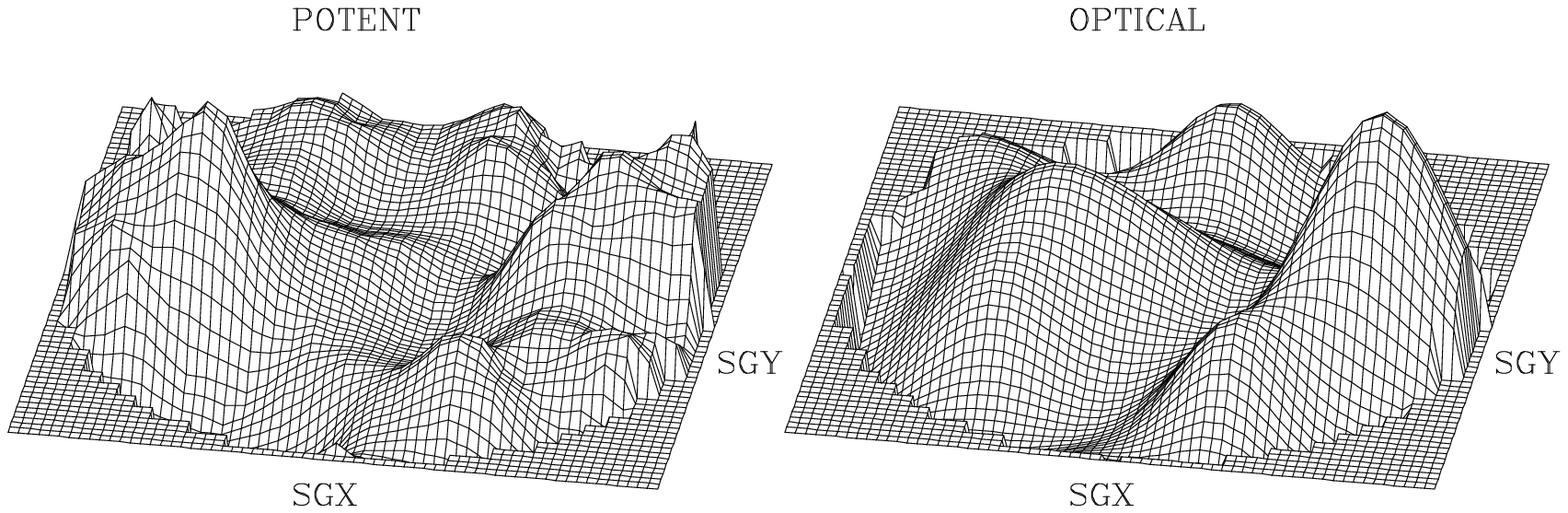}
}
\end{center}
\caption[]{
Surface plots of the density fields in the Supergalactic Plane.  The
left-hand panel shows the \POTENT\ mass density field and the right-hand
panel shows the density field of optical galaxies, both smoothed with
a Gaussian filter of radius 1200 \kms.  The density contrast is
proportional to the height of the surface above (or below) the plane
of the plot.  }
\label{fig:surfmaps}
\end{figure*}

The \POTENT\ density field 
(D95a)
is shown in the left-hand panels of Figs 
\ref{fig:maps} and \ref{fig:surfmaps}. 
\figref{maps} shows contour plots of the smoothed density field in 
slices parallel to the Supergalactic Plane. \figref{surfmaps} shows 
the density field in the Supergalactic Plane as a surface plot where 
the height of the surface above the plane of the plot is proportional 
to density. 

\section{The Optical Density Field} 
\label{sec:opt} 

We now summarize the optical galaxy sample, completeness corrections 
and the method used to reconstruct the density field of optical 
galaxies (for further details see Hudson 1993a). The basic catalogue 
of 12\,439 galaxies covering 67 per cent of the sky is a merger of the 
diameter-limited UGC (Uppsala General Catalogue of Galaxies: Nilson 
1973) and ESO (The ESO-Uppsala Survey of the ESO(B) Atlas: Lauberts 
1982) catalogues. The UGC catalogue is corrected for incompleteness 
for diameters between 1.2 and 1.6 arcmin, and the ESO catalogue is 
complete at diameters larger than 1.4 arcmin (Hudson \& Lynden-Bell 
1991). The 6747 redshifts in the sample were obtained from the major 
redshift surveys, i.e., CfA1 (Davis \& Huchra 1982; Huchra et al.\ 
1983), SSRS (da Costa et al.\ 1988, 1991), SPS (Dressler 1991), and 
Perseus--Pisces (Giovanelli \& Haynes 1985; Giovanelli et al.\ 1986), 
and from the redshift compilations of Huchra (1990) and Fairall \& 
Jones (1991). The variable completeness of the redshift sample is 
mapped as a function of angular diameter and position on the 
sky. These completeness functions are used in conjunction with the 
UGC and ESO diameter functions (Hudson \& Lynden-Bell 1991) to assign 
a weight to each galaxy in the redshift sample to compensate for those 
galaxies in the underlying density field that were not selected. 

We then iteratively transform from redshift space to distance space by 
self-consistently calculating linear peculiar velocities from the 
optical density field (cf.\ Yahil et al.\ 1991). The value of 
$\beta\sbo$ is fixed at 0.5 for these iterations, but the resulting 
optical density field does not depend sensitively on this choice. 
Indeed, had we used $\beta\sbo = 0.75$ in the iteration, our final 
result for $\beta\sbo$ would have increased by only 0.03. 

There are no data in the Zone of Avoidance $|b| < 12\dgr$ and the 
`missing' equatorial strip between the UGC and ESO catalogues 
($-17\ddgr5 < \dec < -2\ddgr5$). We adopt a simple scheme for filling 
in this `masked' area: the `cloned mask' model (cf.\ Lynden-Bell, 
Lahav \& Burstein 1989) attempts a crude interpolation in the galactic 
and equatorial missing strips by copying the set of weighted galaxies 
in adjacent strips and shifting these `clones' in galactic latitude 
and declination respectively. 

In the final step, the weighted set of points is smoothed to yield an 
optical density field. We use the same smoothing filter as used for 
the \POTENT\ density field, namely a Gaussian filter with a 1200 \kms\ 
radius, and we sample the density fields on the same 500 \kms\ grid. 
In the final comparison, we exclude all grid points within the masked 
area (see \secref{vol} below). However, note that, due to the 
smoothing, the density assumed 
within 
the masked region will affect the 
density at adjacent grid points. 
Therefore, the 
`cloning' procedure is preferred over less realistic schemes such as 
the `average mask model' in which 
the masked region 
is set to 
the average density. 

The mean density of optical galaxies is needed in order to calculate 
the density contrast, $\delta\sbo = (\rho\sbo -\overline{\rho\sbo}) 
/\overline{\rho\sbo}$. In this paper, we estimate the mean optical 
density by averaging over optical galaxies in the unmasked volume 
within 6000 \kms. This volume is close to the typical depth of the 
final comparison volume (see \secref{vol} below). 

The optical density field is shown in the right-hand panels of Figs 
\ref{fig:maps} and \ref{fig:surfmaps}.

\section{Determination of $\beta\sbo$} 
\label{sec:comp} 

In the determination of the parameter $\beta\sbo$, our philosophy is 
to carry out as simple an analysis as possible. We use simplifying 
approximations where we believe, based on our experience with mock 
data sets, that any biases introduced by these approximations are 
small compared to the unavoidable uncertainties due to 
the random errors, 
the limited volume sampled (cosmic scatter), 
the unknown complexity of the true biasing relation, 
and the remaining uncorrected effects of non-linear gravity. 
We deviate from a straightforward analysis only when faced 
with a significant bias. The approximations made in this analysis are 
tested using mock catalogues elsewhere (Dekel et al., in preparation,
hereafter D95d). 

In \secref{vol}, we define the volume 
used for the comparison. In \secref{fit}, we discuss the method of 
fit, based on a simple $\chi^2$ statistic, and the effective number of 
independent volumes within the comparison volume. In \secref{sg}, we 
discuss the issue of SG bias and how it biases the determination of 
$\beta\sbo$, and in \secref{scatter} we discuss the scatter in the 
density--density plots. In \secref{results}, we present results for 
$\beta\sbo$ assuming $\om =1$, and in \secref{omega} we discuss the 
results for $\om = 0.3$. 

\subsection{Definition of the comparison volume} 
\label{sec:vol} 

The density--density comparison in this paper is local 
(unlike the velocity--velocity comparison) 
and the \POTENT\ data are weighted inversely by their local errors. 
Consequently, we are free to choose the volume considered in the 
density--density comparison. In selecting the volume, we aim to 
minimize the SG bias and the random errors, and to maximize the number 
of effectively independent volumes used in the comparison. In 
practice, there is an obvious trade-off between these two goals. We 
therefore limit the volumes by $R_4$, in order to minimize the SG 
bias, and by $\sigd$, in order to eliminate regions with very large 
random errors. Furthermore, we use only the grid points that 
are within $7000 \kms$ and outside the masked region of the optical 
density field. In order to investigate how our results depend on 
specific choices for these parameters, we consider two comparison 
volumes for each of the \POTENT\ methods (P0 and P1). We refer to these as 
the S (`small') and L (`large') volumes. \tabref{vol} lists the 
$R_4$ and $\sigd$ limits, the number of grid points (from a cubic grid 
of 500 \kms\ spacing) used in the comparison, and the effective number of 
independent points in the fit (discussed below) for the four volumes 
considered. 

\begin{table}
\caption{Comparison volumes.}
\label{tab:vol}
\begin{tabular}{lcccc}
Sample & $R_4$ & $\sigd$ & $N\sb{grid}$ & $N\sb{eff}$ \\
       & \kms \\
P1 S & $< 1200$ & $< 0.25$ & 2197 & 22.1 \\
P1 L & $< 1500$ & $< 0.30$ & 3334 & 31.2 \\
P0 S & $< 1200$ & $< 0.30$ & 1712 & 18.0 \\
P0 L & $< 1500$ & $< 0.40$ & 2934 & 28.9 \\
\end{tabular}
\end{table}

The limits of the P1L volume are indicated in \figref{maps}. 
The radial indentations in the comparison volume arise from the optical mask 
(i.e.\ the Zone of Avoidance and the `missing' equatorial strip). 
The 
irregular outer limit is determined by 
the inhomogeneity in the \POTENT\ data. 
The P1S volume covers approximately 67 per cent of the sky and 
has a total volume of 
$(6500 \kms)^3$. 
Its 
outer boundary (outside of the Zone of 
Avoidance) lies between 3000 and 7000 \kms, depending on direction. 

\subsection{Method of fit} 
\label{sec:fit} 

A visual comparison between the left- and right-hand panels of 
\figref{maps} and \figref{surfmaps} 
shows a reasonable resemblance between the large-scale features of 
the \POTENT\ and optical density fields. 
Displacements of density peaks are seen at a similar level 
in the mock catalogues 
as a result of the sparse, non-uniform and noisy sampling. 
Assuming GI and no systematic errors, 
the mass density -- optical density diagram is a direct representation of 
the biasing relation 
(for the value of $\om$ used in the \POTENT\ reconstruction). 
Assuming 
a linear biasing relation with no scatter, 
we expect the densities to be related by 
\begin{equation} 
\delta\sb{P} = \lambda \delta\sbo + c 
\end{equation} 
where the slope, $\lambda$, is just equal to $b\sbo^{-1}$. The 
zero-point, $c$, allows for a relative offset in the mean density 
of the two density fields. 
This parameter is necessary because the volumes within which we have 
separately normalized the \POTENT\ and optical density fields are 
slightly different from each other. 

The random errors in the inferred peculiar velocities introduce 
an error in the \POTENT\ density, $\sigd$, so of course we expect 
some scatter in the \POTENT\ -- optical density diagram. 
We have estimated the random shot-noise errors in the optical density 
field using the technique of bootstrap resampling. We find these 
errors to be typically $\sim 0.07$, and always less than half the 
random errors of the \POTENT\ density field at a given grid point. In 
the following analysis we assume that all errors are in the \POTENT\ 
density field, and thus regress $\delta\sb{P}$ on $\delta\sbo$ in the 
determination of $\lambda$. 

If the grid points were independent of each other and if \POTENT\ 
errors were Gaussian, we could construct a reduced $\chi^2$ statistic, 
\begin{equation} 
S = N\sb{grid}^{-1} \sum_i^{N\sb{grid}} 
\frac{(\delta\sb{P} - \lambda \delta\sbo - c)^2} 
{\sigma\sb{P}^2} 
\ , 
\label{eq:reducedchi} 
\end{equation}%
where the sum is over all $N\sb{grid}$ grid points in the 
comparison volume and $\sigma\sb{P}$ is the error in the \POTENT\ density. 

In practice, we sample the density fields on a grid with a much finer 
spacing (500 \kms) than the smoothing radius (1200 \kms), so the grid 
points are not independent. The effective number of independent 
volumes in the $\chi^2$ fit 
can be estimated by 
\begin{equation} 
N\sb{eff}^{-1} = N\sb{grid}^{-2} 
\sum_j^{N\sb{grid}} \sum_i^{N\sb{grid}} 
\exp(-r_{ij}^2/2 R\sb{s}^2) 
\end{equation} 
(DBY93). The values of $N\sb{eff}$ 
for each comparison volume 
are given in \tabref{vol}. We find that with our grid spacing and 
smoothing we have oversampled the volume by a factor $F = 
N\sb{grid}/N\sb{eff} \sim 100$. Thus our data points would be 
independent if sampled on a grid with spacing $F^{1/3} 
\times 500 \kms\ = 2320 \kms$, which is about twice the smoothing 
length. Due to the oversampling, the appropriate statistic to use is 
a corrected $\chi^2\sb{eff} \equiv N\sb{eff} S$, which is distributed 
like $\chi^2$ with $N\sb{d.o.f} = N\sb{eff} - 2$ degrees of freedom. 
This statistic is used to assess the errors on $\lambda$ and $c$, 
as a simple approximation to the elaborate and more accurate 
likelihood analysis using Monte Carlo simulations (e.g. DBY93).

\begin{figure*}
\begin{center}
\epsfysize=12.5cm
\mbox{
\epsfbox{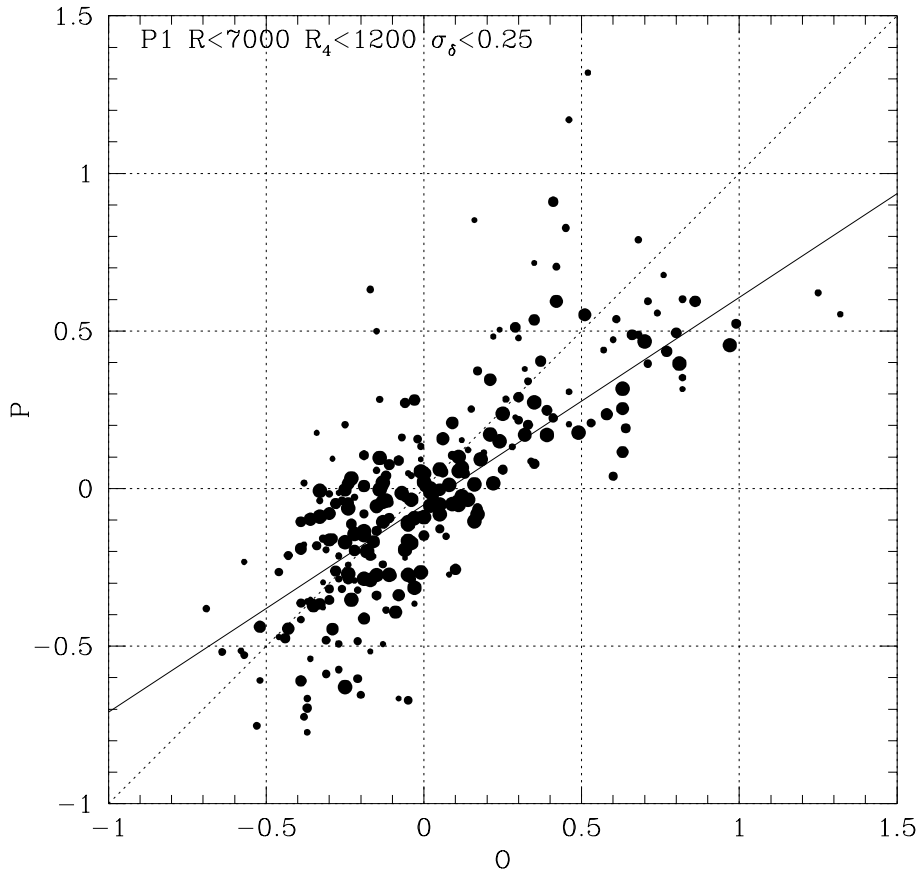}
}
\end{center}
\caption[]{
The density--density scatterplot with the optical density (O) on the
horizontal axis and the \POTENT\ mass density (P) on the vertical axis.
The densities are sampled on a 1000 \kms\ grid within the 
`P1 Small' volume, 
thus the plotted points oversample the number of independent
volumes by a factor $\sim 12$.  The areas of the plotted points have
been scaled with the inverse squares of their errors.  
The solid line is the result of the weighted linear regression  
of $\delta\sb{P}$ on  $\delta\sbo$;      
a dotted line of slope unity is also shown for reference.
}
\label{fig:pvso}
\end{figure*}

\subsection{Correction of biases in \POTENT} 

\label{sec:sg} 
\figref{pvso} shows the comparison between 
$\delta\sb{P}$ (P1S) and $\delta\sbo$. If we set $\sigma\sb{P} = 
\sigma_{\delta}$ (the random error in the \POTENT\ density) in 
\eqref{reducedchi}, we obtain the result $\lambda = 0.66 \pm 0.08$, $c 
= -0.05 \pm 0.02$. The quoted errors are the formal errors from the 
$\chi^2$ fit. This fit is shown as the solid line in \figref{pvso}. 

However, it would be premature to adopt these results as our best 
estimate of $\beta\sbo$ because this estimate is known to be biased. 
DBY93 tested the \POTENT\ method by comparing the smoothed {\it IRAS\/} density 
field to the density field recovered by the \POTENT\ method from the 
unsmoothed {\it IRAS\/}-predicted peculiar velocities with $\om =1$ and $b_I 
=1$. The {\it IRAS\/} velocity and density fields served there as a model for 
a general gravitating system. DBY93 found that, in the limit of dense, 
uniform sampling, they recovered a slope close to the true slope of 
unity with very small scatter (see e.g.\ their fig. 4c). However, 
when they sampled the {\it IRAS\/} peculiar velocity field at only the 
positions of the galaxies in the Mark II peculiar velocity data 
sample, they found an increased scatter and a biased slope of 0.66. 
Thus, the systematic errors in the \POTENT\ procedure are correlated 
with the density field and introduce a bias in the slope, tending to 
reduce it. 

\begin{figure*}
\begin{center}
\epsfysize=18cm
\mbox{
\epsfbox{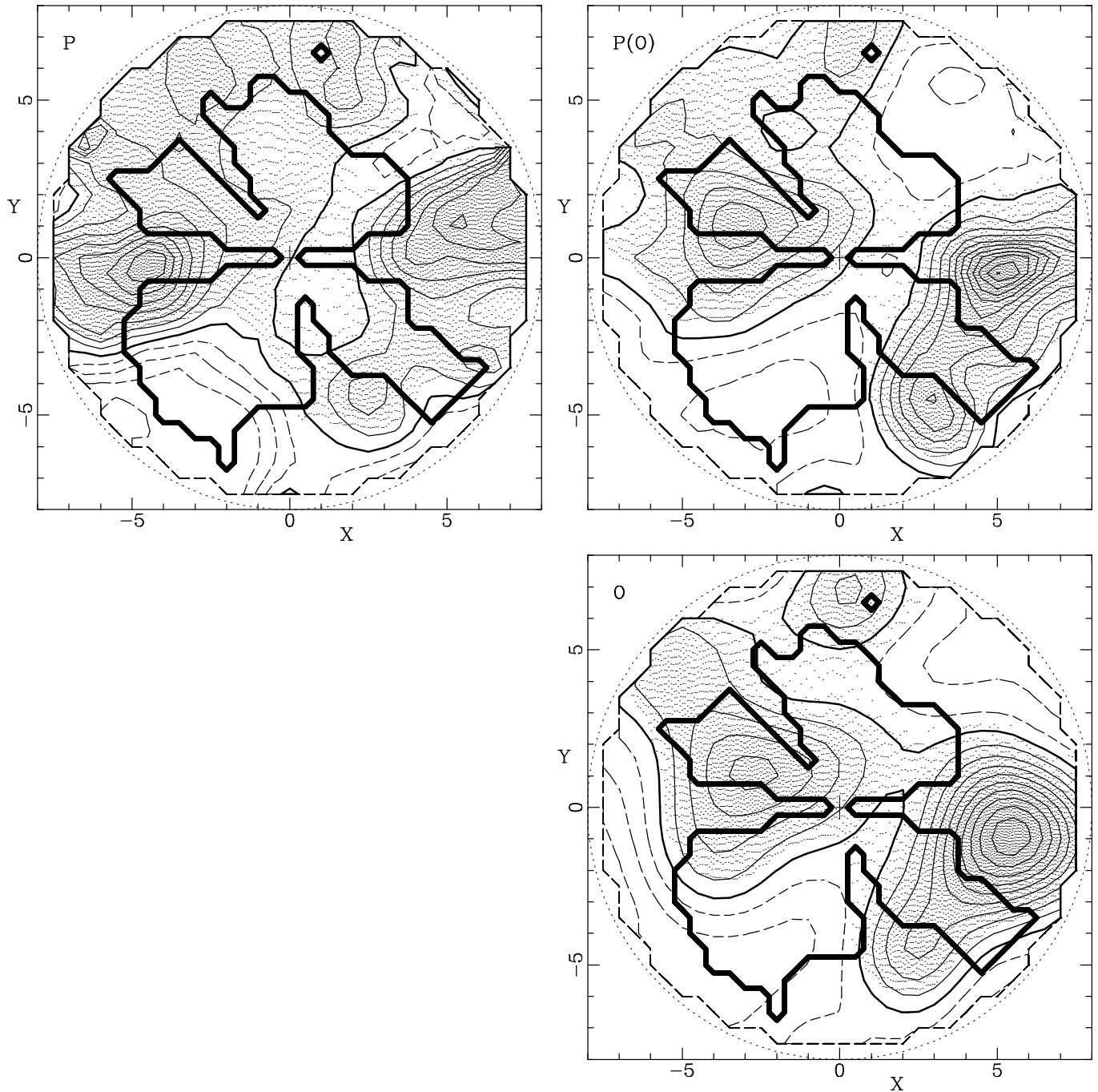}
}
\end{center}
\caption[]{
Maps of the density field in the Supergalactic Plane.  The contour
levels are as in Fig.\ 1.  The upper left-hand panel shows the \POTENT\
mass density field (P). The upper right-hand panel shows P(O), the \POTENT\
recovery using predicted peculiar velocities from the optical density
field at the positions of the Mark III data. The lower right-hand panel
shows the optical density field (O).  The difference between P(O) and
O shows the effect of the biases discussed in the text.
}
\label{fig:ppoomaps}
\end{figure*}

Following DBY93, we use as a test case the optical density field 
itself from which we calculate predicted peculiar velocities via 
gravity at the estimated distance of each galaxy in the peculiar 
velocity data set. Specifically, we begin by smoothing the optical 
density field with a Gaussian filter of 500 \kms\ radius. Peculiar 
velocities are calculated using the prescription of Nusser et al.\ 
(1991), 
i.e. 
\begin{equation} 
\bv{v}(\bv{r}) = \frac{f(\om)}{4 \pi b\sbo} 
\int \frac{\delta\sbo(\bv{r'})} 
{1 + 0.18 b\sbo^{-1}\delta\sbo(\bv{r'})} 
\frac{\bv{r'} - \bv{r}} 
{|\bv{r'} - \bv{r}|^3} \dif^3\bv{r'} 
\, , 
\label{eq:predpv} 
\end{equation}%
at the observed positions of the galaxies in the Mark III data set. 
For the $\om = 1 $ case, we adopt 
a biasing parameter of 
$b\sbo = 1.3$, which ends up consistent with our final result. 
These predicted peculiar velocities are then 
used as the input to the \POTENT\ procedure with $1090\kms$ Gaussian 
smoothing 
which, when added in quadrature with the 
$500 \kms$ smoothing radius of the input field, 
ensures a total Gaussian smoothing of $1200\kms$. 
The resulting mass density fluctuation field is then multiplied by 
$b\sbo$ to obtain the corresponding field of optical galaxies. 
The result of this procedure, which we label P(O), 
is a galaxy density field which is affected by SG bias in the same way 
as the \POTENT\ mass density field recovered from the Mark III data. 
\figref{ppoomaps} compares 
the density fields of \POTENT\ [P], optical through \POTENT\ [P(O)], and 
raw optical [O]. 

\begin{figure*} 
\begin{center}
\epsfysize=12.5cm
\mbox{
\epsfbox{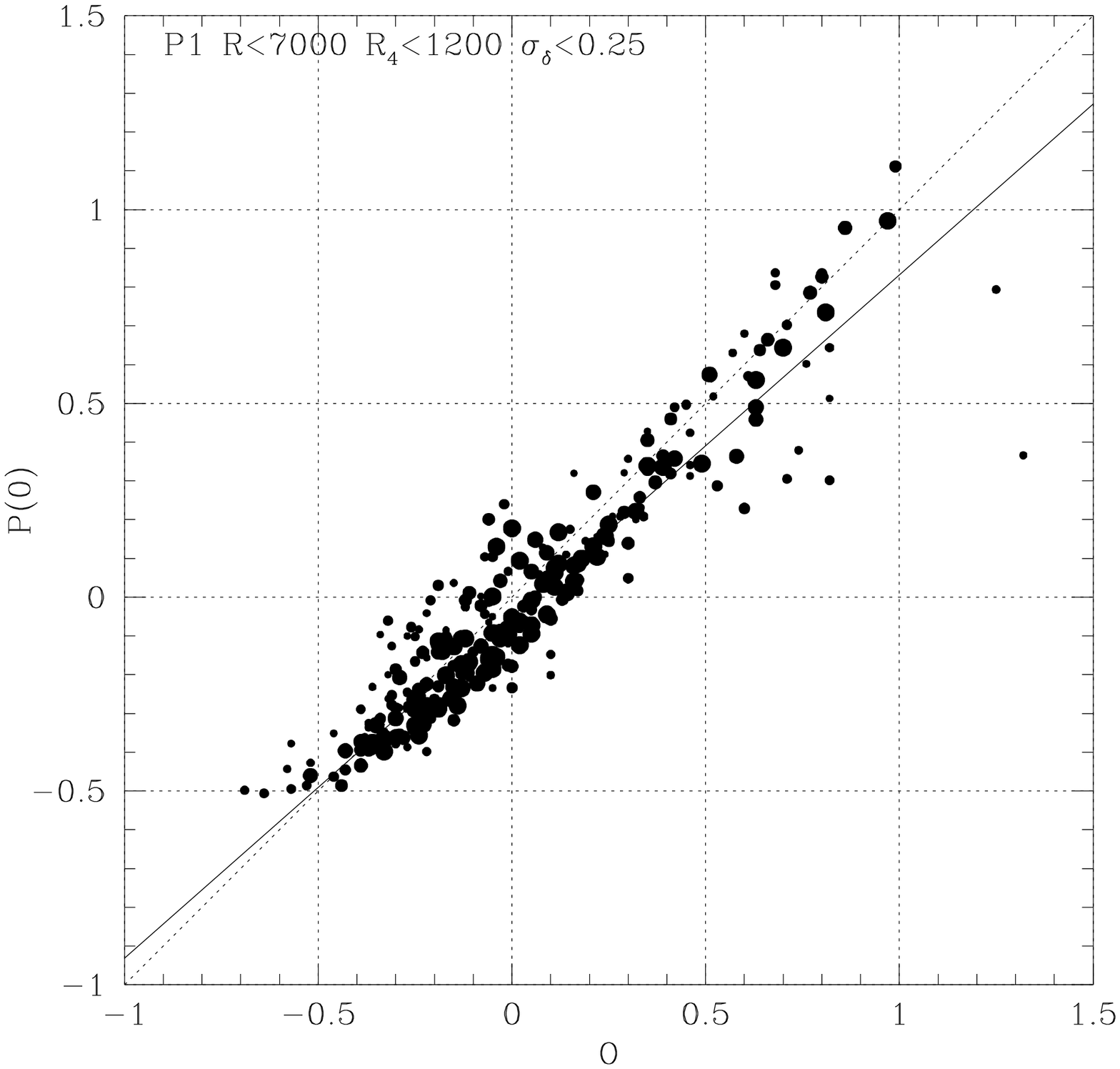}
}
\end{center}
\caption[]{
The density--density scatterplot with optical density (O) plotted on
the horizontal axis and P(O), the \POTENT\ recovery using predicted
peculiar velocities from the optical density field at the positions of
the Mark III data, plotted on the vertical axis.  The comparison
volume, sampling and symbols are as in Fig. 3.  The solid line
is the weighted linear regression of $\delta\sb{P(O)}$ on
$\delta\sbo$; a dotted line of slope unity is also shown for
reference.  The deviation from a slope of unity is due to 
the biases discussed in the text. }
\label{fig:povso}
\end{figure*}

In \figref{povso}, we show the P(O) -- O scatterplot for this test 
case. If the same weights at each grid point as used in the P -- O 
comparison above are adopted, the slope of best fit is 0.88. 
This bias in the slope is smaller than that found by DBY93 due to the 
increase in sampling density of the peculiar velocity data, the 
improvements in the \POTENT\ procedure, and the more accurate adjustment 
of the \POTENT\ smoothing scale to a total of 1200$\kms$. 

\subsection{Scatter in the \POTENT\ -- optical density diagram} 
\label{sec:scatter} 

Before turning to the results of the fits, we must first discuss the 
scatter in the density--density diagram, in order to verify that the 
assumed model of GI plus linear biasing is indeed consistent with the 
data. We find that, by setting $\sigma\sb{P}$ equal to the error in 
the Monte Carlo simulations $\sigd$, we obtain $\chi^2\sb{eff} \gtsim 
2 N\sb{eff}$, which is apparently too large for an acceptable fit. 
This could be a result of underestimating the errors. Indeed, it is 
likely that there are other sources of error, predominantly in the 
POTENT density field, which are not accounted for by the Monte Carlo 
perturbations of the distances. 
We therefore allow for additional errors via an additional parameter, 
$\sigadd$. The additional errors may or may not be correlated with 
$\sigd$, so we assume a weak correlation by adding {\em linearly\/}: 
$\sigma\sb{P} = \sigadd + \sigd$. We then adjust $\sigadd$ so that 
$\chi^2\sb{eff} = N\sb{eff} - 2$. The resulting values are $\sigadd 
=0.06$ for the P1 comparisons, and $\sigadd = 0.04$ for the P0 
comparisons. In \secref{results}, we show that the value of the 
slope, $\lambda$, is not significantly affected by our choice of 
$\sigadd$. It is 
none the less 
necessary to include this additional error 
in order to obtain errors on the slope that are consistent with the 
observed 
scatter in the density--density plot. 

Are these values 
of $\sigadd$ reasonable compared to the possible sources of 
scatter 
that are not included in $\sigd$, and which are not 
eliminated by the comparison of P with P(O)? 
Additional sources of error might include the following. 

\begin{enumerate}

\item{ 
The Monte Carlo estimates, $\sigd$, may underestimate the total errors 
in the \POTENT\ density field. We have made a preliminary test of these 
errors using an ungrouped version of the mock catalogue described 
above. We have compared densities obtained from applying the \POTENT\ 
procedure to the unperturbed velocities of galaxies at their true 
distances and to the perturbed velocities at the perturbed positions, 
after correcting for IM bias. We find that in addition to the Monte 
Carlo errors, $\sigd$, we require $\sigadd=0.03$ in order to obtain an 
acceptable value of $\chi^2$.} 
\item{ 
There may be errors in the \POTENT\ density field due to imperfections in the 
inhomogeneous Malmquist bias corrections. An upper limit to these 
errors is estimated by the rms difference between the \POTENT\ density 
fields corrected and uncorrected for IM, which is $\sim 0.11$.} 
\item{ 
There may be systematic errors in the optical density field due to
interpolation (`cloning') into the unsampled regions. An upper limit
to these errors is estimated by comparing the cloned mask density
field to the average mask density field.  Within the P1S comparison
volume, the rms scatter between the cloned and average mask density
fields is $\sim 0.06$.  Spurious large-scale gradients may also exist
at some level due to possible errors in matching the UGC and ESO
data in the northern and southern hemispheres (see Hudson 1993b,1994b). We
note, however, that the results for $\beta\sbo$ from these two
hemispheres separately are in perfect agreement with each other (see
\secref{results} below).  }
\item{ 
The shot-noise error in the optical density, although smaller than the 
random errors in \POTENT, is not negligible. Within 6000 \kms, the 
typical shot-noise error is $\sim 0.07$. } 
\item{ 
Spatial variations in the efficiency of galaxy formation and 
deviations from a linear biasing relation would generate true, 
physical, scatter about an assumed linear biasing relation between 
galaxies and mass. A scatter of $\sim 0.15$ in the optical galaxy 
density added in quadrature to the \POTENT\ error would account for the 
additional scatter in \figref{pvso}}. 
\end{enumerate} 

Since these 
possible sources of scatter 
are of the same order as our adopted value of 
$\sigadd$, we may conclude that there is no evidence for inconsistency 
between the data and the assumptions of GI and linear biasing, 
allowing us to proceed to the parameter fit. 

\begin{table}
\caption{Results of P -- P(0) fits.}
\label{tab:res}
\begin{tabular}{lr@{$\pm$}lr@{$\pm$}l}
Sample &
\multicolumn{2}{c}{$\lambda$} &
\multicolumn{2}{c}{$c$} \\
\bf P1S & \bf 0.74& \bf 0.13& \bf -0.01& \bf 0.04\\
P1L & 0.78& 0.13&-0.02& 0.04\\
P0S & 0.65& 0.15& 0.01& 0.04\\
P0L & 0.68& 0.14& 0.01& 0.04
\end{tabular}
\end{table}

\subsection{Results} 
\label{sec:results} 

\begin{figure*}
\begin{center}
\epsfysize=12.5cm
\mbox{
\epsfbox{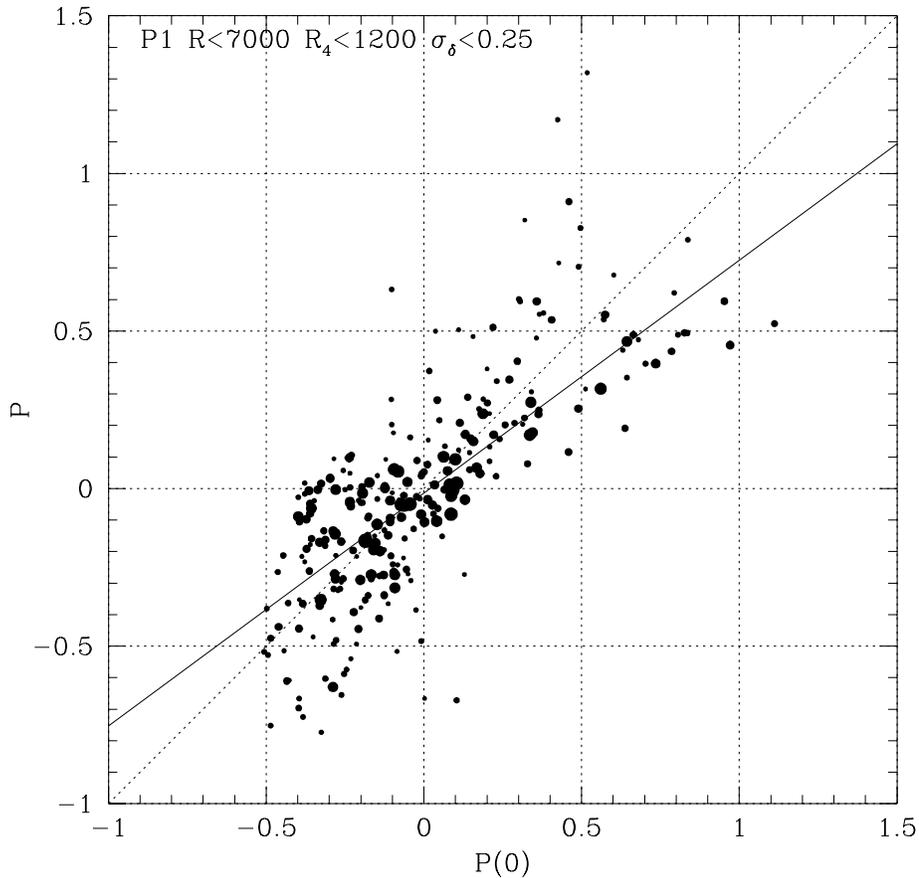}
}
\end{center}
\caption[]{
The density--density scatterplot with P(O) plotted on the horizontal
axis and the \POTENT\ density (P) plotted on the vertical axis.  The
comparison volume, sampling and symbols are as in Fig. 3.  The
solid line is the straight line of best fit; a dotted line of slope
unity is also shown for reference.  The slope $\lambda$ is our 
estimate of $b\sbo^{-1}$ for $\om = 1$.  }
\label{fig:pvspo}
\end{figure*}

The comparison between the \POTENT\ density and the \POTENT-recovered 
optical density P(O) removes the biases in the \POTENT\ procedure and 
thus yields an unbiased estimate of $\beta\sb{O}$. 
As discussed above, we assume that the errors in the \POTENT\ density 
field dominate. We therefore obtain the slope from a regression of 
$\delta\sb{P}$ on $\delta\sb{P(O)}$ with the \POTENT\ error 
$\sigma\sb{P} = \sigma_{\delta} + \sigadd$. The results of the fit 
for the PO and P1 (L and S) cases are summarized in \tabref{res}. 
\figref{pvspo} shows the P -- P(O) scatterplot for the P1S case. 
In all cases, the zero-point $c$ is small and compatible with zero. 
We have made a number of tests to check the robustness of the value of 
$\lambda$ derived from the regression of $\delta\sb{P}$ on 
$\delta\sb{P(O)}$. The results of these tests are summarized below. 

\begin{enumerate} 
\item{ 
The value of $\lambda$ is insensitive to the details of the assumed 
POTENT errors, 
provided 
that the optical errors are negligible. For the P1S case, if we 
assume equal errors for all grid points, our resultant $\lambda$ 
increases by 
only 
3 per cent. 
Similarly, if we set $\sigadd = 0$, 
then 
$\lambda$ decreases by 5 per cent. If we fit by minimizing absolute 
deviation, which is less sensitive to outliers, then $\lambda$ 
decreases by 3 per cent. 
} 
\item{ 
The value of $\lambda$ is insensitive to the exact volume used, as 
indicated by the good agreement between the S and L volumes. A 
further check can be made by dividing the P1S volume into the northern 
celestial hemisphere (containing the Perseus--Pisces supercluster) and 
the southern hemisphere (containing the Great Attractor region). We 
find $\lambda = 0.73 \pm 0.26$ and $0.75 \pm 0.15$ for northern and 
southern hemispheres, respectively. In fact, this almost perfect agreement between 
these independent volumes may indicate that we have overestimated our 
random errors. 
} 
\item{ 
The results are insensitive to the details of the Malmquist 
corrections. As a test of these corrections, we have compared the 
optical density field to a \POTENT\ density field in which the input 
data were corrected for Malmquist bias assuming a homogeneous galaxy 
density field (hereafter HM-corrected).
Although in a few specific regions (e.g. 
Perseus--Pisces) the difference between the IM- and HM-corrected \POTENT\ 
density fields is as large as $\sim 30$ per cent in $\delta\sb{P}$, when 
averaged over all grid points in the comparison volume the net effect 
is small. We find that with the HM corrections $\lambda$ is only 2 per cent 
higher than when IM corrections are used. } 
\item{ 
The values for $\lambda$ are $\sim 12$ per cent lower 
for 
the P0 case than 
for 
the P1 case. This is a crude measure of the typical systematic error 
associated with the \POTENT\ procedure. } 
\item{ 
We have compared the P1S case to a version of the optical density 
field in which the masked region was set to the average density before 
smoothing (the `average mask model' from Hudson 1993a). The value of 
$\lambda$ is $\sim 15$  per cent higher than the result with the `cloned 
mask' density field used here. Although the average mask model is 
unrealistic, it sets an upper limit on the systematic error arising 
from the treatment of the masked region. } 
\end{enumerate} 

Our estimate of $\beta\sbo$, adopting the P1S results of the 
$\delta\sb{P}$ on $\delta\sb{P(O)}$ regression, is $\beta\sbo = 
0.74\pm0.13$, where the quoted 1$\sigma$ errors are the formal 
errors on the slope which are consistent with the scatter about the 
linear fit. 

Note that if we assume that the `extra' scatter in the 
density--density diagram lies in the optical density field, whether 
due to errors in the optical density field or due to a real physical 
effect, such as spatial variations in the efficiency of galaxy 
formation, then we must do a more elaborate regression which allows 
for errors in both the \POTENT\ and optical density fields. As an 
example, if we set $\sigadd = 0$ and thus $\sigma\sb{P} = \sigd$, but 
allow for a 0.15 error in the optical density, the regression yields a 
slope $\lambda = 1.00 \pm 0.17$ and an acceptable value of 
$\chi^2\sb{eff}$. 
In \secref{discuss} we discuss the possibility that there may be 
significant scatter in the efficiency of galaxy formation.

\subsection{The dependence on $\om$} 
\label{sec:omega} 

The \POTENT\ reconstruction includes quasi-linear effects so, in 
principle, it 
should 
be possible to obtain some information on $\om$, 
independent of $b\sbo$. To test this idea, we have generated a \POTENT\ 
reconstruction with $\om$ set to the value 0.3. This is compared to a 
P(O) density field in which the predicted peculiar velocities are 
determined via \eqref{predpv} with $\om=0.3$, and $b\sbo$ = 0.7 in the 
quasi-linear correction within the integral. The slope of the 
resulting fit is thus an estimate of $b\sbo^{-1}$. 

We find a slope $\lambda = b\sbo^{-1} = 1.63\pm0.33$, corresponding to 
$b\sbo = 0.61\pm0.12$ or $\beta\sbo = 0.79\pm0.16$. Thus, we recover 
essentially the same value of $\beta\sbo$ as in the $\om=1$ case 
because we are only in the mildly non-linear regime. The scatter 
about the straight-line fit in the $\om=0.3$ case is marginally 
smaller than in the $\om =1$ case, but the improvement is not 
statistically significant. We conclude that, given the random and 
systematic errors, including the possible deviations from linear 
biasing, we cannot remove the degeneracy between $\om$ and $b\sbo$. 
In order to rule out a low-density universe by these results, one has 
to appeal to external arguments such as the fact that the required 
degree of {\em anti-biasing\/} is physically implausible.

\section{Discussion and Summary} 
\label{sec:discuss} 

Our simple analysis does not allow a formal evaluation of goodness of 
fit between the data and the assumed model of GI and linear biasing, 
but we conclude, based on the expected random and systematic errors 
and the possible real scatter in the biasing relation, 
that the data and model are not in conflict. 
Note, however, that a non-gravitational origin to the peculiar 
velocities, which could invalidate the current analysis, 
cannot be ruled out. 
An 
apparent 
agreement 
between the \POTENT\ density field and the galaxy density field can also 
occur in non-gravitational scenarios provided that 
the galaxy density field has developed from homogeneous initial 
conditions in a way that satisfies the continuity 
equation, and that the present-day velocity field is irrotational and 
proportional to the time-averaged velocity field 
(Babul et al.\ 1994). 

Adopting the model of GI and linear biasing, our main goal has been to 
determine $\beta\sbo$ on a scale of 1200 \kms. We find, from a 
regression of \POTENT\ density on optical density, 
the result 
$\beta\sbo = 0.74 \pm 
0.13$ (formal $1\sigma$ error). This result is robust to the details 
of the volume used, 
to the treatment of 
Malmquist bias, and to the actual value of $\om$. 
The regression of \POTENT\ density on optical density is 
insensitive to the relative weighting of grid points in the fit. 
The sensitivity of $\beta\sbo$ to extreme changes in the \POTENT\ 
weighting scheme is at the 12 per cent level. 
Its sensitivity to extreme changes in the way in which the optical 
density field is interpolated across of the Zone of Avoidance is at a 
similar level. 

Non-linear effects on this scale are small compared to the 
uncertainties in the data, the analysis, and the biasing scheme, so we 
are unable to remove the degeneracy between $\om$ and $b\sbo$. Our 
degenerate result corresponds to $b\sbo=1.35\pm0.23$ if $\om=1$, and 
to $\om = 0.61\pm0.18$ if $b\sbo=1$. Low values of the density 
parameter ($\om \ltsim 0.3$) would require optical galaxies to be 
significantly anti-biased with respect to mass. 

One should 
be cautioned not to interpret the estimated value of 
$\beta\sbo$ too literally. On top of the formal error of 0.13 and the 
systematic uncertainties of similar order, our estimate of $\beta\sbo$ 
is subject to cosmic scatter in the mean density; the comparison 
volume of radius $\ltsim 6000 \kms$ may still be small for a `fair' 
sample. This cosmic scatter arises because, in our analysis, both the 
peculiar velocity data input to \POTENT\ {\em and\/} the mean density of 
optical galaxies are normalized within a volume of radius $\sim 6000$ 
\kms. Note that while the free parameter $c$ corrects for small 
differences in the normalizing volumes of the \POTENT\ and optical 
density fields, and consequent differences in the mean density of 
these fields with respect to each other, it does not necessarily 
correct to the universal density. Thus our result is a {\em local\/} 
value of $\om$ measured within the comparison volume. The rms scatter 
of the mean density in such a volume is expected to be at the level of 
$\sim 10-15$ per cent according to the fluctuation power spectra in 
conventional cosmological scenarios. 
Thus the local value of $\om$ may differ from the universal 
value at a similar level. 

We have assumed throughout that the additional scatter in the 
density--density scatterplots (over and above that due to the distance 
errors which scatter the \POTENT\ densities) is due to other small 
unquantified 
errors in the \POTENT\ 
density fields. As 
discussed in \secref{scatter}, the amount of additional scatter is 
comparable to our estimates of these systematic errors. However, it 
is also possible that spatial variations in the biasing relation 
between galaxies and mass may be large on scales of 12$h^{-1}$ Mpc. 
For example, 
Cen \& Ostriker (1993) 
investigated hydrodynamic simulations of 
the Cold Dark Matter cosmology 
and 
found on scales of 10$h^{-1}$ Mpc a non-linear biasing 
relation which corresponds roughly to $b\sbo\simeq 1.2 - 1.4$ in the range 
$-0.5 < \delta < 1.0$, in good agreement with our results. The rms 
scatter in their biasing relation is about 0.23 in $\delta\sbo$ at 
$\delta\sbo = 0$, comparable to the scatter of $\sim 0.15$ which we 
would need to add in quadrature to the optical galaxy density to 
obtain good fits. If there are significant spatial variations in the 
efficiency of galaxy formation, then this adds an astrophysical source 
of scatter in 
the $\delta$ -- $\delta\sbo$ plot 
which must be accounted for in the analysis. 
If this extra scatter is assigned to
$\delta\sbo$ then the 
slope of the regression changes to $\beta\sbo \sim 1$. 

The current estimates of $\om$ and $b$ from large-scale structure are 
reviewed by Dekel (1994, section 8 and his table 1). In particular, the result 
for $\beta\sbo$ found here is lower than the value $\beta_I = 
1.28^{+0.75}_{-0.59}$ (95 per cent confidence) found by DBY93, who compared 
an earlier realization of \POTENT\ with the 1.9-Jy {\it IRAS\/} density field. 
A comparison between the 1.2-Jy {\it IRAS\/} density field and the new version 
of \POTENT\ is in progress (D95d). 
If we allow for the fact that {\it IRAS\/} galaxies are less clustered 
(biased) than optical galaxies by a relative factor of $\sim 1.3$ on 
the scales studied in this paper (Hudson 1993a; Peacock \& Dodds 1994), 
then the different results for $\beta$ lead to compatible values of 
$\om$. 

Hudson (1994b) compared predicted peculiar velocities in linear theory 
from the optical density field smoothed with a 500 \kms\ top-hat 
filter to observed peculiar velocities from the Mark II catalogue, and 
obtained the result $\beta\sbo = 0.50\pm 0.06$ (where the quoted 
errors are formal 1$\sigma$ random errors only). It seems at this 
early 
stage, 
while 
the various analyses are still in progress and 
the systematic effects are not fully understood, that 
velocity--velocity comparisons tend in general to yield somewhat lower 
estimates for $\beta$ than the estimates based on density--density 
comparisons 
(e.g. Nusser \& Davis, 
private communication; 
Willick et al., in preparation). 
However, the results are not at all incompatible when one allows for 
the systematic uncertainties in both determinations. A possible 
physical explanation for this difference in $\beta$ is a variation in 
the biasing relation with scale or density. 

The lower bounds on $\om$ from cosmic flows independent of biasing 
(e.g. Nusser \& Dekel 1993; Dekel \& Rees 1994; see the review by Dekel 
1994) are currently at the level of $\om >0.3$ with 99 per cent confidence, 
consistent with the theoretically favoured value of $\om=1$. 
Other input of relevance include 
(i) constraints from the microwave background fluctuations, 
(ii) rms number fluctuations of optical and {\it IRAS\/} 
galaxies in top-hat spheres of radius $8\hmpc$ 
($\sigma\sb{8,O}\approx 0.95$, $\sigma_{8,I}\approx 0.65$), 
and 
(iii) theoretical expectations for biasing 
based on first-generation simulations with gas dynamics. 
A 
choice of parameters that
is consistent with 
the result 
of this paper and all the constraints above 
would have mass fluctuations at the level of $\sigma_8\approx 0.75$, 
and linear biasing factors $b\sbo\approx 1.3$ and $b_I\approx 0.9$ 
for optical and {\it IRAS\/} galaxies respectively. 

\section*{Acknowledgments} 
This research has been supported by PPARC through the rolling grant 
for Extragalactic Astronomy and Cosmology at Durham, the US-Israel 
Binational Science Foundation, and the Israel Science Foundation. 
MJH and AD acknowledge the hospitality of the Institut d'Astrophysique 
de Paris where part of this work was done. 

 \end{document}